\documentclass[a4paper,aps,preprintnumbers,twocolumn,amsmath,amssymb,prl]{revtex4}
\usepackage{bbm}
\usepackage{mathrsfs}
\usepackage{amsmath}
\usepackage[dvips]{graphicx}
\usepackage{epsfig}
\usepackage[dvips]{graphics,graphicx}

\newcommand{\be}{\begin{equation}}
\newcommand{\ee}{\end{equation}}
\newcommand{\bea}{\begin{eqnarray}}
\newcommand{\eea}{\end{eqnarray}}
\newcommand{\bes}{\begin{split}}
\newcommand{\ees}{\end{split}}

\usepackage[dvips]{graphics,graphicx}

\begin{document}
\title{Effect of the lattice alignment on
Bloch oscillations of a Bose-Einstein condensate in a square optical lattice}
\author{ M.-C. Chung$^1$ and A. R. Kolovsky$^{2,3}$}
\affiliation{$^1$Max-Planck-Institut f\"ur Physik komplexer Systeme,
01187 Dresden, Germany}
\affiliation{$^2$Kirensky Institute of Physics, 660036 Krasnoyarsk, Russia}
\affiliation{$^3$ Siberian Federal University, 660041 Krasnoyarsk, Russia}

\begin{abstract}
We consider a Bose-Einstein condensate of ultracold atoms loaded
into a square optical lattice and subject to a static force. For
vanishing atom-atom interactions the atoms  perform periodic Bloch
oscillations for arbitrary direction of the force. We study the
stability of these oscillations for non-vanishing interactions,
which is shown to depend on an alignment of the force vector with
respect to the lattice crystallographic axes. If the force is
aligned along any of the axes, the mean field approach can be used
to identify the stability conditions. On the contrary, for a
misaligned force one has to employ the microscopic approach, which
predicts periodic modulation of Bloch oscillations in the limit of a
large forcing.
\end{abstract}

\date{\today}
\maketitle

A Bose-Einstein condensate (BEC) in optical lattices has intrigued a
rapidly growing interest as it provides an experimentally realizable
system with controllable interactions. A variety of phenomena
concerning different aspects of physics for quantum many-body
systems has been studied, such as the superfluid-Mott insulator quantum
phase transition \cite{Greiner,Stoeferle}, BEC-BCS crossover for
fermionic gases \cite{Bartenstein}, and quantum transport in
accelerated lattices, where atoms exhibit fundamental quantum
effects such as the Wannier-Stark ladder \cite{Raizen}, Landau-Zener
tunnelling \cite{Arimondo}, and Bloch oscillations
\cite{Raizen,Morsch} -- phenomena usually associated with
electron in solid crystall. This work deals with the last mentioned
problem, namely, Bloch oscillations (BO) of condensed atoms in
optical lattices. We would like to mention that besides pure
academic interest this problem also has an applied aspect because BO
provide a tool for precision  measurement of gravitational field and
inter-atomic interaction constant.

Until quite recently almost all theoretical, numerical and
experimental studies of BO concerned 1D or quasi 1D lattices (see
Ref.~\cite{63,Oberthaler,Konotop} for the contemporary reviews).
Nowadays one observes a growing interest in BO in multidimensional
lattices \cite{51_55_58,Witthaut,Snoek,Trompeter,Gustav,Widera}. In
the single-particle approach this problem was considered in
Refs.~\cite{51_55_58,Witthaut}. It was shown that an increase of the
lattice dimensionality introduces new effects not present in the 1D
lattice. Some predictions of these works were later on confirmed in
the experiment with the array of optical guides \cite{Trompeter},
where one uses a formal analogy between the Maxwell and
Schr\"odinger equations. The experiment with a BEC of interacting
atoms addresses the further questions \cite{Widera}, in particular,
the question about the stability of multidimensional BO. Indeed, it
is known that a BEC in optical lattices can be dynamically unstable,
which quantum-mechanically means decoherence of the BEC \cite{Eva}.
In the present work we study the conditions under which the
dynamical instability is suppressed and, hence, multidimensional BO
are stable. Unlike 1D lattices, these conditions are shown to
involve an alignment of the static force vector with respect to the
crystallographic axes of the lattice. We also argue in the work that
by changing the angle between the primary lattice vectors and the
force vector one may observe a transition from the mean-field to the
microscopic Bloch dynamics.

To simplify the equations we shall consider  the two-dimensional
case throughout the paper, - generalization of the results in three
dimensions is straightforward. The Bose-Hubbard Hamiltonian of atoms
in the tilted 2D lattice reads,
\begin{equation}
\label{1}
 \begin{array}{ll}
\widehat{H}= & -\frac{J_x}{2}\sum_{m,l} \left(
\hat{a}^\dag_{m+1,l}\hat{a}_{m,l}+h.c.\right) \\
& -\frac{J_y}{2}\sum_{m,l} \left(
\hat{a}^\dag_{m,l+1}\hat{a}_{m,l}+h.c.\right) \\
& +\frac{W}{2}\sum_{m,l} \hat{n}_{m,l}(\hat{n}_{m,l}-1) \\
& + d\sum_{m,l}(F_x m+F_y l)\hat{n}_{m,l} \;,\\
 \end{array}
\end{equation}
where $J_{x,y}$ are the hopping matrix elements in $x$ and $y$
directions, $W$ microscopic atom-atom interaction constant, $d$
lattice period, and $F_{x,y}$ the projections of the static force
vector on the lattice axes. The Hilbert space of (\ref{1}) is
spanned by the Fock states $|{\bf n}\rangle\equiv|n_{m,l}\rangle$,
where $\sum_{m,l} n_{m,l}=N$ -- the total number of atoms. Since in
the coordinate representation the Fock states are given by the
symmetrized product of the localized Wannier functions, we shall
refer to this basis as the Wannier basis. The translational
invariance of the system, broken by the static term, can be actually
recovered by using the gauge transformation \cite{63}. Then the
Hamiltonian (\ref{1}) takes the form
\begin{equation}
\label{2}
 \begin{split}
\widehat{H}(t) = & -\frac{J_x}{2}\sum_{m,l} \left( e^{-i\omega_x t}
\hat{a}^\dag_{m+1,l}\hat{a}_{m,l}+h.c.\right)\\
& - \frac{J_y}{2}\sum_{m,l} \left(e^{-i\omega_y t}
\hat{a}^\dag_{m,l+1}\hat{a}_{m,l}+h.c.\right) \\
& +\frac{W}{2}\sum_{m,l} \hat{n}_{m,l}(\hat{n}_{m,l}-1) \;,
 \end{split}
\end{equation}
where $\omega_{x,y}=dF_{x,y}/\hbar$ are the Bloch frequencies
associated with $x$ and $y$ component of the static force. We also
note that in stead of the Wannier basis one can use the
quasimomentum Fock basis $|{\bf q}\rangle\equiv|q_{p,k}\rangle$
for the Hamiltonian (\ref{2}), which we shall refer to as the
Bloch basis. (Needless to say that in the coordinate representation
the quasimomentum Fock states are given by the symmetrized product
of the extended Bloch functions.) Formally this corresponds to the
canonical transformation
\begin{displaymath}
\hat{b}_{p,k}=\frac{1}{L}\sum_{m,l}
\exp\left[-i\frac{2\pi}{L}(mp+kl)\right] \hat{a}_{m,l} \;,
\end{displaymath}
which implicitly assumes the periodic boundary conditions.

\begin{figure}
\center
\includegraphics[width=8.5cm]{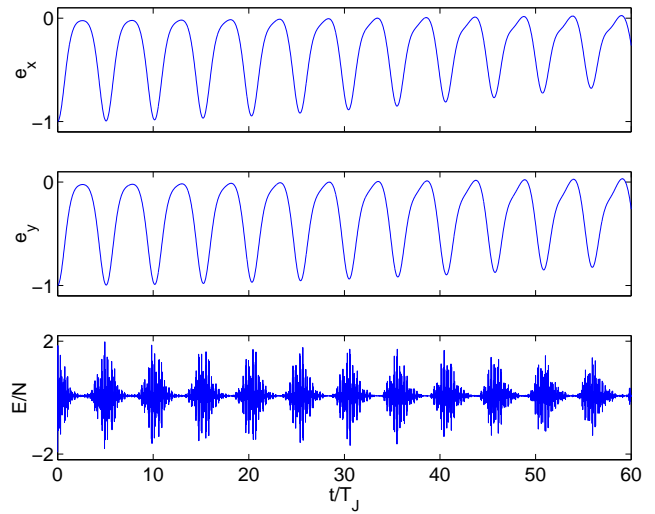}
\caption{Bloch oscillations of condensed atoms in the 2D lattice: the order
parameter $e_x(t)$ (top), $e_y(t)$ (middle), and the mean energy
(bottom). The system parameters are $N=7$, $L=3$ (periodic
boundary conditions), $J_{x}=J_{y}=J$, $W=0.2J$, $Fd=20J$, and
${\bf F}/F=(\sqrt{2/5},\sqrt{3/5})$. The time is measured in units
of the tunnelling period $T_J=2\pi\hbar/J$.} \label{fig1}
\end{figure}

\bigskip
{\em Misaligned Force}---We begin with the case of a strong
misaligned force $dF_x, dF_y \gg J_{x,y}>W$. In order to illuminate
situation, we model BO in a small 2D lattice by numerically solving
the time-dependent Schr\"odinger equation with the Hamiltonian
(\ref{2}) for the specified initial conditions. As those we consider
the superfluid state with $q_{p,k}=N\delta_{p,0}\delta_{k,0}$ ,
which approximates the ground state of the system for ${\bf  F}=0$
and $W<J_{x,y}$. (Substitution of this state by the exact ground
state practically does not affect the final result.) Fig.~\ref{fig1}
shows the numerical results for a $3\times 3$ lattice with $7$ atoms
inside. The lower panel in Fig.~\ref{fig1} depicts the mean energy
of the system, the upper and middle panels show the order parameters
$e_x(t)$ and $e_y(t)$ defined as \cite{Smerzi}
\begin{equation}
\label{10} e_x(t)=  -\frac{1}{N}{\rm Re}\left[\langle\Psi(t)|
\sum_{m,l} \hat{a}^\dag_{m+1,l}\hat{a}_{m,l}|\Psi(t)\rangle\right]
\;.
\end{equation}
[By replacing the operators in the bracket in Eq. (\ref{10}) with
$\sum \hat{a}^\dag_{m,l+1}\hat{a}_{m,l}$ one obtains a similar
expression for the order parameter $e_y(t)$.] It is seen in the
figure that BO persist in time but are modulated with some characteristic
period. We would also like to mention that the Bloch dynamics
displayed in Fig.~\ref{fig1} is converged in the thermodynamic
limit, i.e., for given $\bar{n}=N/L^2$ the further increase of the
system size affects neither the modulation period nor the shape of
modulation.

To prove that BO in the misaligned lattice are stable in
the limit of strong forcing and to identify the modulation period
we proceed as follows.  First we introduce the new wave function
$|\widetilde{\Psi}(t)\rangle$ through the relation
$|\Psi(t)\rangle=\widehat{U}_0(t)|\widetilde{\Psi}(t)\rangle$,
where $\widehat{U}_0(t)$ is the evolution operator
for vanishing atom-atom interactions. The function
$|\widetilde{\Psi}(t)\rangle$ obviously obeys the equation,
\begin{equation}
\label{6}
i\hbar\frac{\partial|\widetilde{\Psi}(t)\rangle}{\partial t}
=\frac{W}{2}\widehat{U}_0^\dag(t)\left( \sum_{m,l}
\hat{n}_{m,l}(\hat{n}_{m,l}-1)\right)
\widehat{U}_0(t)|\widetilde{\Psi}(t)\rangle \;.
\end{equation}
On the other hand, the explicit form of the evolution operator is
given by $\widehat{U}_0(t)=\widehat{T}^\dag \widehat{D}(t)
\widehat{T}$, where the unitary operator $\widehat{T}$ represents
the transformation from the Wannier basis $|{\bf n}\rangle$ to the
Bloch basis $|{\bf q}\rangle$ and the matrix of the operator
$\widehat{D}(t)$ is diagonal in the Bloch basis,
\begin{equation}
\label{8}
 \begin{split}
\langle {\bf q}|D(t)|{\bf q}\rangle =
& \exp\left[i\frac{J_x}{dF_x}\sum_{i=1}^N \sin\left(
\frac{2\pi p_i}{L}-\omega_x t\right) \right. \\
& \left. +i\frac{J_y}{dF_y}\sum_{i=1}^N
\sin\left(\frac{2\pi k_i}{L}-\omega_y t\right) \right] \;.
 \end{split}
\end{equation}
Note that the operator (\ref{8}) tends to the identity
operator for $F_x,F_y\rightarrow\infty$. Substituting
$\widehat{U}_0(t)$ in Eq.~(\ref{6}) by identity matrix and noting
that the interaction energy operator is diagonal in the Wannier
basis with integer entries, $\langle {\bf n}|\sum_{m,l}
\hat{n}_{m,l}(\hat{n}_{m,l}-1)|{\bf n}\rangle =\sum_{m,l} n^2_{m,l}-
N$, we conclude that the time evolution of the wave function
$|\widetilde{\Psi}(t)\rangle$ is periodic with the period
$T_W=2\pi\hbar/W$ \cite{nature}. Coming back to the original wave
function this result means the periodic modulation of BO with the
frequency $\omega_W=W/\hbar$. It is worth stressing that the above
proof assumes both $F_x$ and $F_y$ to be large and, hence, the case
of aligned lattices is excluded.

\bigskip
{\em Aligned Force}---Next we consider the situation where the force
is aligned along one of the crystallographic axes (to be certain,
the y-axis in what follows). Within the single-particle approach the
static force $F_y$ would localize the atoms in the $y$-direction.
Thus one may expect that if $F_y$ is large enough the atoms form
separate BECs in the planes perpendicular to the force vector,
weakly coupled together as a one-dimensional BEC chain. Introducing
new operators  $\hat{A}_{l} = \frac{1}{\sqrt{L}}\sum_{m}
\hat{a}_{m,l}$ and $\hat{A}^\dag_{l}$, the effective Hamiltonian
reads
\begin{equation}
\label{12}
\begin{split}
\widehat{H}_{eff}=
& -J_x\sum_l \hat{A}^\dag_{l}\hat{A}_l \\
& -\frac{J_y}{2}\sum_l \left(
e^{-i\omega_y t} \hat{A}^\dag_{l+1}\hat{A}_l+h.c.\right) \\
& +\frac{W_{eff}}{2}\sum_l \hat{N}_l(\hat{N}_l-1) \;,
 \end{split}
\end{equation}
where $W_{eff}=W/L$. Thus we have reduced the 2D problem to a 1D
problem with the renormalized interaction constant. (If one
considers 3D lattices, the renormalization is $W_{eff}=W/L^2$.)
Moreover, since the mean number of atoms $\widetilde{N}$ in any
site of the effective 1D system is given by $\bar{n}L$, the
occupation numbers will be macroscopically
large in the thermodynamic limit $N,L\rightarrow\infty$,
$\bar{n}=N/L^2=const$, which justifies the mean field approach.

The mean-field Hamiltonian of the system (\ref{12}) reads (up to the
irrelevant constant terms proportional to $\sum_l \widetilde{N}_l =
N$)
\begin{equation}
\label{13}
H_{eff}=-\frac{J_y}{2}\sum_l \left( e^{-i\omega_y t}
A^*_{l+1}A_l+h.c.\right) +\frac{g}{2}\sum_l |A_l|^4 \;,
\end{equation}
where $A_l$ and $A_l^*$ are pairs of the canonically conjugated
variables and the macroscopic interaction constant
$g=W_{eff}\widetilde{N}=W\bar{n}$. Within the mean-field approach
the border between stable and unstable (decaying) BO is know
exactly \cite{Andrey2,Zheng}. Namely, for $J/Fd>0.5$ the critical
value of nonlinearity is a linear function of the static force
magnitude, while for $J/Fd<0.5$ it additionally depends on the
value of the hopping matrix elements:
\begin{equation}
\label{14}
g_{cr}\approx\left\{
\begin{array}{ll}
0.33Fd \;, & Fd<2J \\ 0.1(Fd)^2/J \;, & Fd>2J
\end{array}\right. \;.
\end{equation}
Obviously, for a fixed nonlinearity $g$ the condition (\ref{14}) can
also be formulated as a condition on the critical magnitude $F_{cr}$
of the static force.
\begin{figure}
\center
\includegraphics[width=8.5cm]{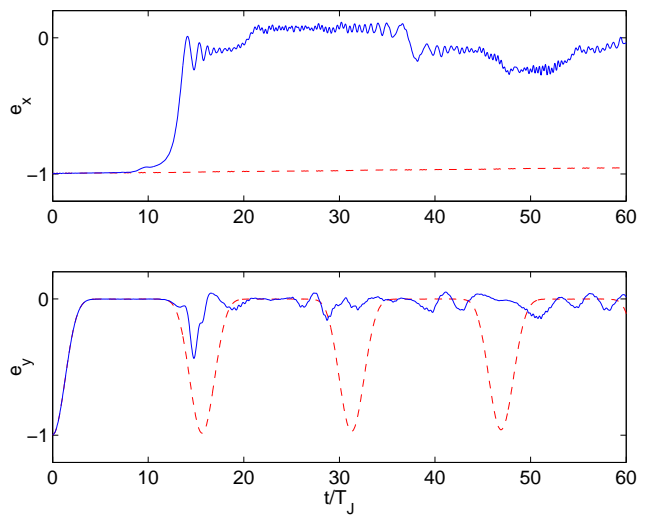}
\caption{Dynamics of the order parameters for ${\bf F}/F=(0,1)$,
dashed lines, and ${\bf F}/F=(0.001,\sqrt{1-0.001})$, solid
lines. The other parameters are the same as in Fig.~1.}
\label{fig3}
\end{figure}

The microscopic analysis of BO in the aligned lattice confirms our
working hypothesis. Choosing the parameters in such a way that the
1D mean-field BO are stable, we simulate BO of $N=7$ atoms in the
2D lattice with $L=3$. The dashed line in the upper panel of
Fig.~\ref{fig3} shows the dynamics of the order parameter $e_x(t)$.
It is seen that $e_x(t)\approx-1$, -- therefore we indeed have
in-plane BECs. We also note that the decay and
revival of the order parameter $e_y(t)$ in the lower panel is
an artifact due to the finite size of our lattice. Indeed, it can
be shown that the time evolution of $e_y(t)$, calculated on the
basis of the effective Hamiltonian (\ref{12}), obeys the equation
\cite{57}
\begin{equation}
\label{15}
e_y(t)= -\exp\left(-2\widetilde{N}\left[1-\cos\left(
\frac{W_{eff} t}{\hbar}\right)\right]\right) \;.
\end{equation}
Because $W_{eff}=W/L$ and $\widetilde{N}=\bar{n}L$, one has
$e_y(t)=-1$ in the thermodynamic limit.

\begin{figure}
\center
\includegraphics[width=8.5cm]{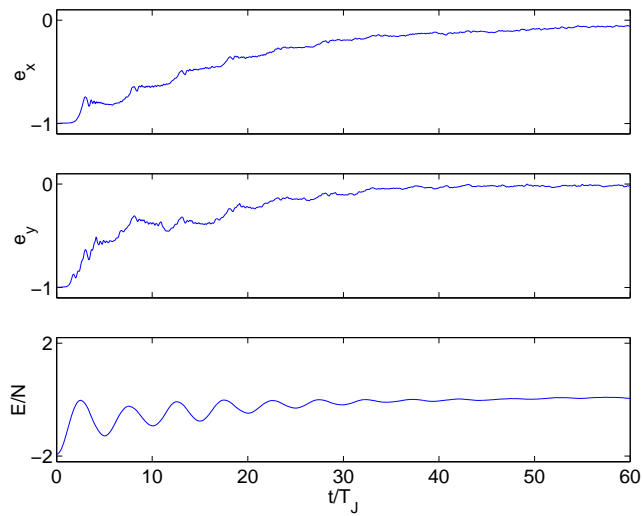}
\caption{Bloch dynamics for $dF=0.2J$ and ${\bf F}/F=(0,1)$. The
other parameters are the same as in Fig.~1.} \label{fig4}
\end{figure}

The above analysis of BO in the aligned lattice relies on the
reduction of  a two-dimensional system to an effective one-dimensional
mean-field problem. It should be especially stressed that this reduction
is possible only if the dynamics of the reduced system is stable.
If we choose  the parameters in the unstable regime
the situation becomes totally different. Figure \ref{fig4} shows the
numerical results for $F = 0.2J/d<F_{cr}$, where one-dimensional BO suffer
from dynamical instability. Unlike in the stable
regime, BO along $y$ direction now excite the transverse degree of
freedom and we observe decay of the {\em both} order parameters towards
zero. Thus no reduction to one dimension is possible.

\bigskip
{\em Slightly Misaligned Force}---Finally we briefly analyze an
experimentally important situation of a small mismatch between the
lattice axis and the static field vector, i.e., $F_x  \ll F_y$. The
solid lines in Fig.~\ref{fig3} show the order parameters for the
same $dF=20J$ but $F_x=0.001F$. Compared to the case $F_x=0$ (dashed
lines in Fig.~\ref{fig3}), we observe the destruction of BEC after
time $t^*\approx 12T_J$. This critical time can be understood in
terms of the mean-field approach as well. Indeed, it is known that a
stationary BEC is unstable for the quasimomentum ${\bf \kappa}$
outside the first quarter of the Brillouine zone. Since the static
force causes the linear growth of the quasimomentum,
$\kappa_{x,y}(t)=\kappa_{x,y} + F_{x,y}t/\hbar$, the system always
enter the instability region of the Brillouine zone. However, if the
static force is strong enough, the system passes the instability region
so quickly that it `has no time' to decay. [In
fact this is a physical argument behind Eq.~(\ref{14}).] In the
considered example the strong static force ensures the fast driving
along $y$ direction but simultaneously it slowly brings the system
to $\kappa_x=\pi/2d$ along $x$ direction. As soon as this border of
instability is reached ($t^*/T_J=J/4dF_x$), we observe an
irreversible decay of the order parameters.

\bigskip
{\em Conclusion.} In summary, we have studied BO of a BEC of atoms
in a square lattice for both aligned and misaligned static forces. It
is shown that in the case of aligned force the system may be reduced
to a one-dimensional chain of mini BECs, which we treat by using the
mean-field approach. Then the stability diagram of this effective 1D
system defines the critical magnitude of the static force above
which BO are stable, with no excitations of the transverse degrees of
freedom. On the contrary, in the unstable regime, $F<F_{cr}$, BO
induced by the static force excite the transverse modes and, as a
consequence, one observes BEC distruction and decay of BO. Our
studies also illuminated importance of the alignment. The
strong ($F>F_{cr}$) but slightly misaligned force is shown to slowly
intrigue the transverse modes, which destabilize BO after some
well-defined transient time. However, if misalignment is large,
BO appear to be stable again. This case corresponds to the quantum
(not mean-field) regime of BO, where they are modulated with the
frequency defined by the microscopic interaction constant.

\bigskip
Fruitful discussions with A.
Buchleitner and financial support by Deutsche Forshungsgemeinschaft
under SPP1116 are gratefully acknowledged.


\end{document}